\documentclass[prb,aps,superscriptaddress,twocolumn,floatfix]{revtex4-2}

\usepackage{graphicx} 
\usepackage{epsfig}
\usepackage{amsmath}
\usepackage{amssymb}
\usepackage{amsfonts}
\usepackage{mathptmx}
\usepackage{dcolumn}
\usepackage{eucal}
\usepackage{bm}
\usepackage{color}
\usepackage{xcolor}
\usepackage[colorlinks,linkcolor=blue,citecolor=blue,allcolors=blue]{hyperref}
\usepackage{url}
\usepackage{bm} 
\usepackage{dsfont}

\newcommand*\imag{\mathrm{\textit{i}}} 
\newcommand*\transp{\mathrm{T}} 

\newcommand*\TRSF{\mathop{}\!\mathcal{T}} 
\newcommand*\PHSF{\mathop{}\!\mathcal{C}} 
\newcommand*\SHSF{\mathop{}\!\mathcal{S}} 
\newcommand*\CC{\mathop{}\!K} 
\DeclareMathOperator{\sgn}{sign} 
\newcommand*\ham{\mathop{}\!\hat{H}} 

\def\real{\Re}
\def\imaginary{\Im}

\begin{document}


\title{Topological phase transitions in
superconductors with chiral symmetry}

\author{Kristian Løvås Svalland}
\affiliation{Center for Quantum Spintronics, Department of Physics, Norwegian
University of Science and Technology, NO-7491 Trondheim, Norway}

\author{Maria Teresa Mercaldo}
\affiliation{Dipartimento di Fisica ``E. R. Caianiello", Universit\`a di Salerno, IT-84084 Fisciano (SA), Italy}


\author{Mario Cuoco}
\affiliation{SPIN-CNR, IT-84084 Fisciano (SA), Italy, c/o Universit\`a di Salerno, IT-84084 Fisciano (SA), Italy}

\begin{abstract}
We study topological transitions in one dimensional superconductors that can harbor multiple edge Majorana bound states
protected by chiral symmetry. The chiral symmetry arises due to the structure of the internal spin degrees of freedom of the superconductor and it can be guided by the coupling of the superconductor with sources of time-reversal symmetry breaking. We then consider distinct regions of the phase diagram in the parameters space that are marked by gapless excitations in the spectrum and evaluate the conditions for inducing a topological transition. We show that for gapless chiral symmetric superconductors one can identify a class of physical perturbations that enable a gap opening in the spectrum, without breaking chirality, and turn the system into a topological state. This type of superconductor is dubbed marginal topological superconductor because an infinitesimally small perturbation is able to induce a transition into a topological nontrivial phase.
To explicitly demonstrate and evaluate the character of the transitions from gapless to topological gapfull phases we explore different physical cases including $p$-wave superconductor in the presence of an applied magnetic field or 
proximity-coupled to a ferromagnet, 
and $s$-wave superconductor in a noncollinear magnetic ordering. 
\end{abstract}

\maketitle

\section{Introduction}

Topological superconductors keep being at the center of intense investigation since the seminal works on nontrivial topology in superfluid helium 3 (He-3)~\cite{Volovik_2009} and on paired states with breaking of parity and time-reversal symmetry~\cite{Read_2000}. Topological superconductors~\cite{Kitaev_2001,Qi2011,Sato_2017,Tanaka2012} are indeed expected to play a key role for the design of innovative quantum transport properties and for the achievement of fault-tolerant quantum computation~\cite{Kitaev_2001,Ivanov2001,Sato2003,Nayak2008,Sau2010,Alicea2011} by exploiting the Majorana modes~\cite{Kitaev_2001,Ivanov2001,Sato2003} they can host. 

Apart from materials that can exhibit an intrinsic topological behavior, there are several quantum platforms that have been proposed to engineer topological superconductivity~\cite{Alicea_2012,Beenakker2013,Sato2016} by achieving a synthetic $p$-wave pairing, as for instance in conventional superconductors 
proximity-coupled to topological insulators~\cite{Fu2008}, semiconductor-superconductor hybrids~\cite{Sau2010,Alicea2010,Lutchyn2010,Oreg2010}, magnetic atom chains on superconductors~\cite{Choy2011,Nadjperge2013,Nakosai2013,Braunecker2013,Klinovaja2013,Vazifeh2013,Pientka2013,Poyhonen2014,Heimes2014,Brydon2015,Li2014,Heimes_2015,Xiao_2015,Andolina2017}, {unconventional superconductors} \cite{Tanaka2024}, and others. 
According to the Altland–Zirnbauer symmetry classification scheme~\cite{Altland1997}, topological superconductors can be related to different symmetry classes. In the above mentioned engineered systems the resulting topological phase belong to the so-called D class, that breaks time-reversal symmetry and in one-dimensional superconducting nanostructure is marked by a Majorana mode occurring at each end of the system. 
Here, in this manuscript we focus on topological superconductors that are part of the BDI class, which fulfills a time-reversal-like symmetry and particle-hole symmetry, and can be characterized by means of an integer topological invariant. 
This type of topological superconductors can host an integer number of gapless Majorana
modes on their boundaries if they are fully gapped in the bulk. In general multi Majorana bound states (MBS) become accessible in both intrinsic ~\cite{Dumitrescu2013,Dumitrescu2014,Mercaldo2016,Mercaldo2018,Mercaldo2018_b,Mercaldo2019}, or
effective ~\cite{Tewari2012,Volpez2019,Haim2014,Klinovaja2014,Kotetes2015,Hell2017,Pientka2017,Kotetes2019,Heimes_2015,Xiao_2015,Andolina2017,Spanslatt2017,Brzezicki2018} $p$-wave spin-triplet superconductors, with the
chiral symmetry being associated with a sublattice symmetry or an internal symmetry of the Hamiltonian. Remarkably, the presence of multiple MBS 
can lead to the design of novel Josephson effects including synthetic Weyl points with chiral anomalies~\cite{Kotetes2019,Mercaldo2019}, topologically protected nodal Andreev spectra ~\cite{Sakurai2020} as well as nonstandard odd-frequency pair correlations~\cite{Takagi2022,Cayao2024odd}.   

Any time a topological phase transition occurs in a system, effectively changing the number of topological modes at the boundary, the gap in the bulk must be vanishing. If the Hamiltonian stays gapless, the topological invariant is in general not defined, but any perturbation which opens the gap again, without altering the symmetry properties, may simultaneously turn the system into a nontrivial topological phase. In principle there are no general rules to engineer physical perturbations that guarantee the changeover from a gapless to a topological phase. 
In this paper we focus on superconductors that owe a chiral symmetry and belong to the BDI class. We then consider distinct regions of the phase diagram in the parameters space where the Hamiltonian is marked by gapless excitations.
For the BDI class we introduce the concept of marginal topological superconductor if the occurrence of gapless excitations of the Hamiltonian is due to a realness criterion associated with the trajectory that sets out the winding number of the system in momentum space (see also Appendix \ref{app:A}). 
For this specific superconducting state, any perturbation which adds an imaginary part to the trajectory that builds up the winding number can in general open the gap and thus bring the system into a topological phase.
Such types of superconducting systems are interesting since they can set out physically relevant paths to obtain and manipulate topological systems by means of arbitrarily small perturbations. In addition, they are potentially easy to obtain since their gaplessness is not dependent on fine-tuning in the phase space.
We study this physical scenario for different types of superconductors in the presence of various sources of time-reversal symmetry breaking in such a way that they preserve the chiral symmetry and keep the superconductor into the BDI class.
In particular, to explicitly demonstrate the realization of this type of dubbed marginal topological phase we investigate the following physical cases: i) p-wave superconductor in the presence of an applied magnetic field or proximised to a ferromagnet, 
ii) s-wave superconductor in the presence of a noncollinear magnetic ordering. 

The paper is organized as follows: In Sec. \ref{sec:sec2_pwave} we  describe the magnetic field driven marginal topological phases for spinfull p-wave superconductors. 
In Sec. \ref{sec:3}, we discuss the case of topological phase transitions with s-wave pairing and spatially modulated exchange field. 
Our conclusions are reported in Sec. IV.

\section[p-wave and FM ordering]{Magnetic field driven marginal topological p-wave superconductivity}
\label{sec:sec2_pwave}

In this Section we introduce the model describing a one-dimensional p-wave superconductor in the presence of an external magnetic field or coupled to a ferromagnet with uniform magnetization. Then, we evaluate the topological number and the resulting phase diagram. We conclude the section by providing the conditions for having topological phase transitions from gapless to gapfull configurations by means of small perturbations that can always results into a topological nontrivial phase.

\subsection{Model and phase diagram}
We consider the Hamiltonian of a {one-dimensional} p-wave superconductor in the presence of a magnetic field or equivalently a uniform exchange field arising from the proximity with a ferromagnet. The Hamiltonian in real space is expressed as
\begin{equation}
    \begin{split}
        \ham =& 
        \sum_{ij\sigma} W(i-j)c_{i\sigma}^\dagger c_{j\sigma} 
        -\sum_{iss'}(\mu\sigma_0 + \bm h\cdot \bm \sigma)_{ss'} c_{is}^\dagger c_{i s'}\\
        &+ \sum_{i\sigma\sigma'} \left[\Delta_{\sigma\sigma'}c_{i\sigma}^\dagger c_{i+1,\sigma'}^\dagger + \mathrm{h.c.} \right].
        \label{eq:fm-pwave-latticeham}
    \end{split}
\end{equation}
with $c_{i\sigma}^\dagger$ the creation operator of an electron with spin $\sigma=\uparrow,\downarrow$ at the site $i$.
Here, \(W(i-j) = W(j-i)\) is a real parameter describing the hopping strength between lattice sites \(i\) and \(j\), \(\bm h\) is the magnetic  field strength, \(\mu\) is the chemical potential and \(\Delta_{\sigma\sigma'} = \Delta_{\sigma'\sigma}\) are the spin triplet superconducting order parameters (OPs). 
In the case of periodic boundary conditions, the Hamiltonian can be expressed in momentum space as 
\begin{equation}
    \begin{split}
        \ham =& \sum_{k\sigma} \epsilon_k c_{k\sigma}^\dagger c_{k\sigma} - \sum_{kss'} c_{ks}^\dagger \left(\bm h \cdot \bm \sigma\right)_{ss'} c_{ks'}\\
        &+\frac{1}{2}\sum_{k\sigma\sigma'}\left(2\imag\sin(k)\Delta_{\sigma\sigma'}c_{k\sigma}^\dagger c_{-k\sigma'}^\dagger + \mathrm{h.c.}\right),
    \end{split}
\end{equation}
where \(\epsilon_k\) is the Fourier transform of \(W(i) - \mu\), especially we define \(W(i) = -\delta_{1,|i|}t - \delta_{2,|i|}t' - \delta_{3,|i|}t''\), where \(t\), \(t'\) and \(t''\) are the nearest neighbour, 
next nearest neighbour, 
and the third nearest neighbour hopping strength, respectively. Thus \(\epsilon_k\) is
\begin{equation*}
    \epsilon_k = -2t\cos(k) - 2t'\cos(2k) - 2t''\cos(3k) - \mu.
\end{equation*}
This Hamiltonian can be cast in tensor form by defining the Nambu spinor 
\begin{equation*}
    \psi_k = (c_{k\uparrow},\ c_{k\downarrow},\ c_{-k\downarrow},\ -c_{-k\uparrow})^\transp,
\end{equation*}
which gives
\begin{align}
    \ham &= \frac{1}{2}\sum_{k} \psi_k^\dagger h_k \psi_k\\
    h_k &= \tau_z \epsilon_k - \bm h \cdot \bm \sigma + (\bm d_k \tau_{+} + \bm d_k^* \tau_{-})\cdot \bm \sigma.
    \label{eq:hk-fm-pwave-tensor}
\end{align}
with \(\bm d_k = 2\sin(k)\bm d\). 
Here the superconducting OPs are defined in terms of the \(d\)-vector as \(\Delta_{\uparrow\uparrow,\downarrow\downarrow} = \pm \imag d_x + d_y\) and \(\Delta_{\uparrow\downarrow} = -\imag d_z\).
Provided that \(h_x=0\) and \(d_x=-d_x^*,\, d_y=d_y^*,\, h_z=h_z^*\), the Hamiltonian satisfies the following time-reversal ($\mathcal{T}$), particle-hole ($\mathcal{C}$) and chiral ($\mathcal{S}$) symmetries 
\begin{align*}
    \mathcal{T} &= \tau_z\sigma_z\mathcal{K}
    &\mathcal{C} &= \tau_y\sigma_y\mathcal{K}
    &\mathcal{S} &= \tau_x\sigma_x \,.
\end{align*}
Furthermore, the operators associated with the particle-hole and time-reversal symmetries are such that their square is equal to \(+1\). 
In the following we assume that the magnetic field lies in the $yz$ plane and it is parameterized by the angle $\theta$ such as $\mathbf{h}=(0,\sin \theta, \cos \theta)$.
Thus,  this Hamiltonian belongs to the BDI class, and in 1D we expect it to have an integer \(\mathbb{Z}\) topological number~\cite{Altland1997}. Such topological number can be also related to a winding number. 
{We review here one way to obtaining it, i.e. by rotating}
the Hamiltonian into the basis of the eigenvectors of the chiral symmetry operator, that is
\begin{align*}
    H \to H' &= U^\dagger H U & U^\dagger \SHSF U = \mathrm{diag}(-1, -1, 1, 1).
\end{align*}
In this way, we get a block off-diagonal matrix
\begin{equation}
\label{eq:block-off}
    H = \begin{pmatrix} 0 & A_k\\ A_{k}^\dagger & 0\end{pmatrix},
\end{equation}
with
\begin{equation}
\label{eq:ak-fm-pwave}
    A_k = \begin{pmatrix}
        2\imag \Delta_{\uparrow\uparrow}\sin(k) - \epsilon_k + h_z & -\imag h_y\\
        -\imag h_y & 2\imag \Delta_{\downarrow\downarrow}\sin(k) + \epsilon_k + h_z
    \end{pmatrix}
\end{equation}

\begin{figure*}
\begin{center}
    \includegraphics[width=0.80\textwidth]{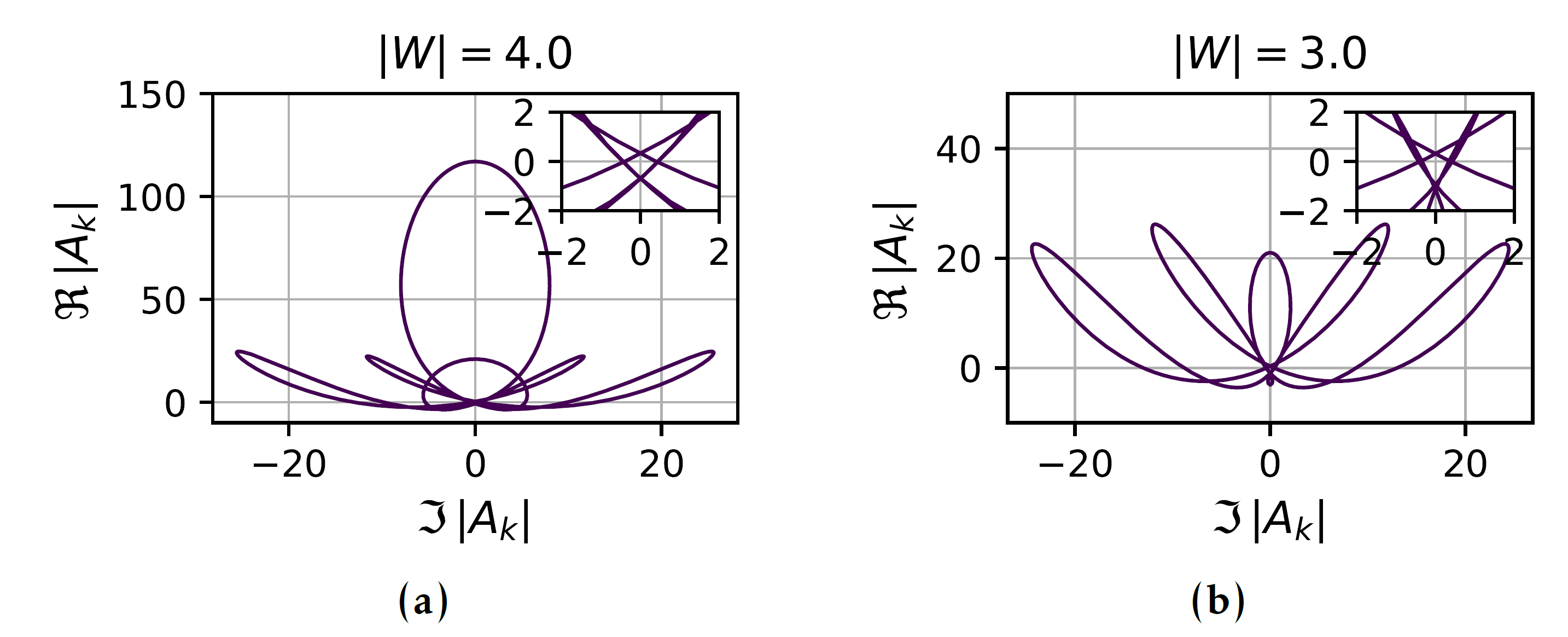}
    \caption{Parametric plots of the determinant $|A_k|$ 
    The common parameters used in the plots are: $\Delta_{\uparrow\uparrow}=2.2t$, $\Delta_{\downarrow\downarrow}=0.18t$, $\mu=1.0t$, $t'=t$; $h=2.0t$ and $\theta=\pi/4$. In panel (a) the third nearest neighbour hopping strength is set to $t''=3t$, while in (b) it is $t''=-2t$, giving different winding numbers, namely $|W|=4$ and $|W|=3$, respectively. }
    \label{fig:wind-curves-fm}
\end{center}
\end{figure*}
{The winding number is given by the  number of times the determinant of $A_k$ winds around the origin in the complex plane as the parameter \(k\) traverses the first Brillouin zone. Hence, defining $z_k=\exp(i\theta_k)=\det A_k/|\det A_k|$, we can write
\begin{equation}
    W = \frac{1}{2\pi\imag}\int \frac{d z_k}{z_k} .
    \label{eq:winding-number-complex}
\end{equation}
} 
In contrast with the 1D spinless Kitaev model on the chain, the winding number in this model can vary between -4 and 4 \cite{Mercaldo2016,Mercaldo2018}, and a phase change can be accompanied by a difference in winding number being
\begin{itemize}
    \item even, for band gap closing at \(k\neq -k\), or
    \item odd or even, for energy gap closing at \(k=-k\).
\end{itemize}
The latter case is in general odd, but can be even in the case that \(k=0\) and \(k=\pi\) closing gaps happen simultaneously. 
Time-reversal and chiral symmetry dictate that when the Pfaffian of the matrix at \(k\) changes sign, so too does the Pfaffian at \(-k\). 
That is, in the BdG formalism with chiral symmetry all energy crossings at \(k\neq -k\) happens in groups of 4. 

What is interesting to note about the last item in the list above is that the \(\hat{\Delta}\) contribution at time reversal invariant points is zero because it has odd-parity. Thus, for any set of the other parameters, no amount of change in the superconducting OPs can change the parity of the winding number. Examples of two phases with different winding number parity is shown in Fig.~\ref{fig:wind-curves-fm}. 
The only difference between the two panels is the value of \(t''\), 
which can change the parity of the winding number. 

As the absolute value of the winding number encodes the number of MBS at each edge, the parity of the winding number is an important value as it decides whether there are single MBS at the edges or if all can be paired in Majorana doublets. If some local symmetry breaking perturbation at the edges couple Majoranas to each other, they will split in energy. The difference between an odd and even parity winding number therefore decides if there still exist an unbound  {Majorana mode} at the edge. If the number of MBS is odd, then there will be at least one Majorana mode left on each edge, whereas in the case of an even number of MBS, no such statement can be given. This can be deduced from the fact that any skew-symmetric \(N\times N\) has a Pfaffian of 0 when \(N\) is odd. Since the effective perturbation Hamiltonian between the Majoranas at the edge has \(N = |W|\), there exist at least one unbound Majorana after a symmetry breaking perturbation is added.

\begin{figure*}[tb]
    \centering
        \includegraphics[width=0.98\textwidth]{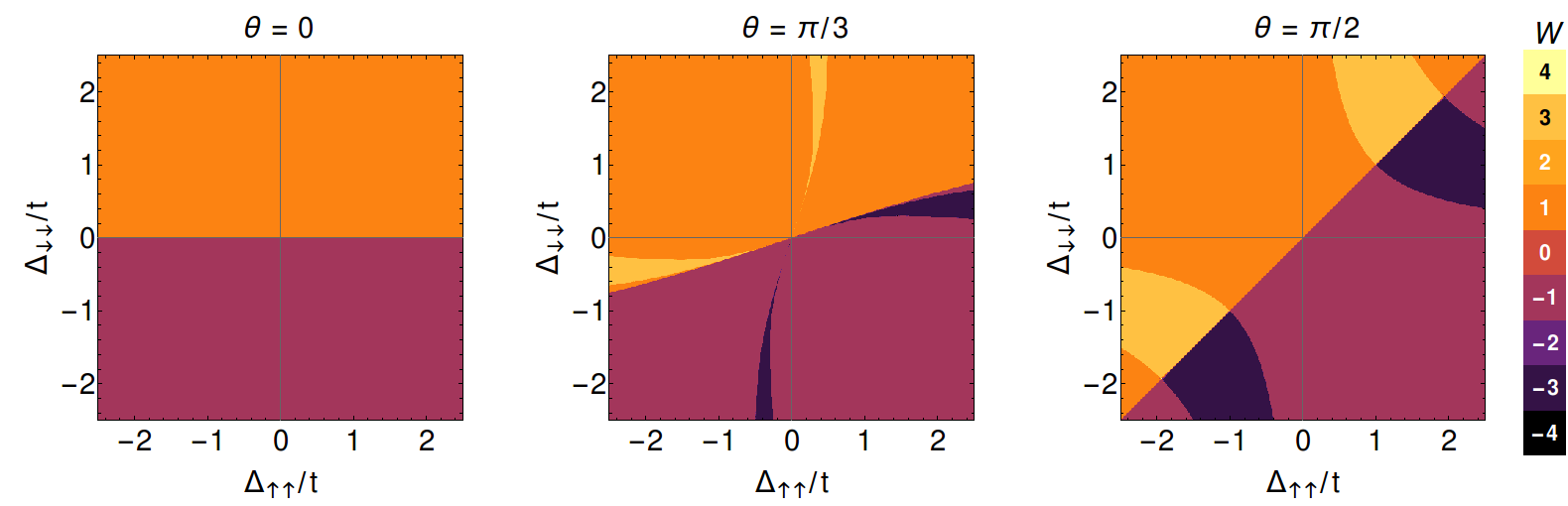}
    \caption[Phase diagram in $(\Delta_{\uparrow\uparrow,\downarrow\downarrow})$-space with odd winding number.]{Phase diagram in $(\Delta_{\uparrow\uparrow,\downarrow\downarrow})$-space for \(\mu = t' = t\), \(t''=-2t\), \(h = 2t\) and \(\theta=0\), \(\theta=\pi/3\) \(\theta=\pi/2\) for left, middle and right panel respectively. $\theta$ is the angle formed by the magnetic field with respect to the $z$-axis. {For this set of parameters W is odd} and, as explained in the text, the parameters \(\Delta\) are not able to change the parity of the winding number because their contribution vanish at \(k=0,\,\pi\).}
    \label{fig:pwave-fm-phasedeltauudd-h2mu2t11-2}
\end{figure*}

\begin{figure*}[tb]
    \centering
     \includegraphics[width=0.98\textwidth]{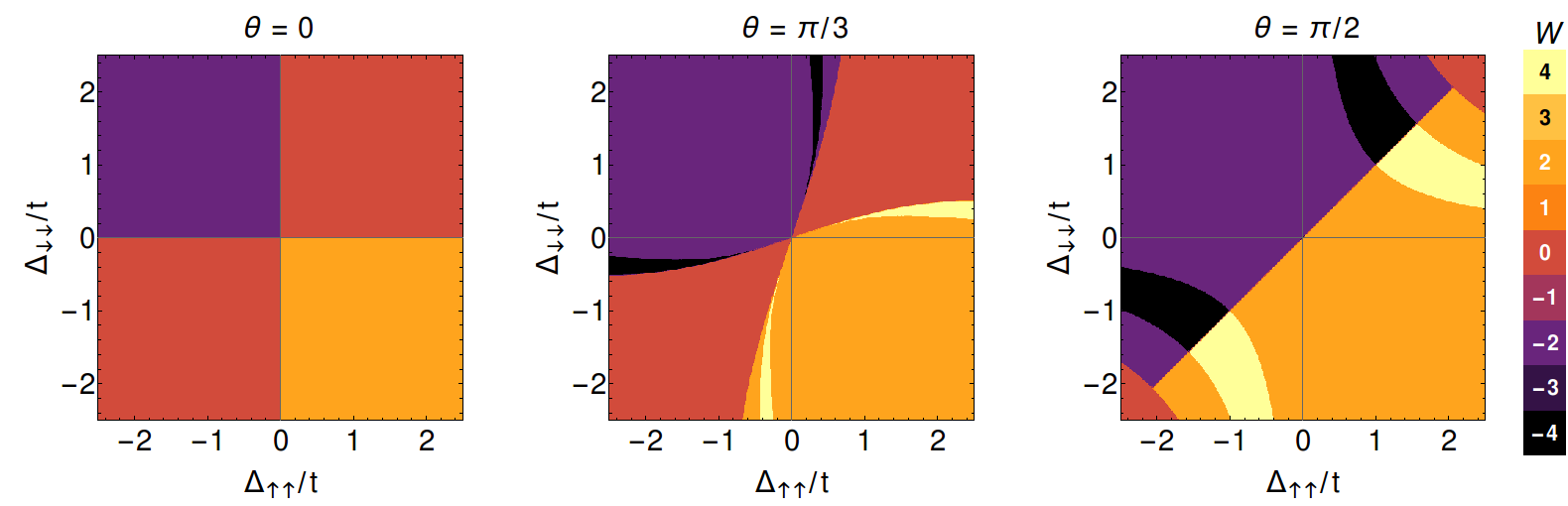}
    \caption[Phase diagram in $(\Delta_{\uparrow\uparrow,\downarrow\downarrow})$-space with even winding number.]{Phase diagram in $(\Delta_{\uparrow\uparrow,\downarrow\downarrow})$-space for \(\mu = t' = t\), \(t''=3t\), \(h = 2t\) and \(\theta=0\), \(\theta=\pi/3\) \(\theta=\pi/2\) for left, middle and right panel respectively. {For this set of parameters W is even} and, as explained in the text, the parameters \(\Delta\) are not able to change the parity of the winding number because their contribution vanish at \(k=0,\,\pi\).}
    \label{fig:pwave-fm-phasedeltauudd-h2mu2t113}
\end{figure*}
The resulting phase diagram defined by the winding number is quite structured, being a function of the parameters in the Hamiltonian. The phase diagram in $(\Delta_{\uparrow\uparrow},\Delta_{\downarrow\downarrow})$-space is given for two representative sets of parameters in Fig.~\ref{fig:pwave-fm-phasedeltauudd-h2mu2t11-2} and Fig.~\ref{fig:pwave-fm-phasedeltauudd-h2mu2t113}.
The results refer to various orientations of the magnetic field and different amplitudes of the second nearest neighbor hopping $t{''}$.
The analysis has been performed to highlight the possibility of engineering solely odd or even values of the winding number in the $(\Delta_{\uparrow\uparrow},\Delta_{\downarrow\downarrow})$-space. Indeed, in Fig.~\ref{fig:pwave-fm-phasedeltauudd-h2mu2t11-2} starting from the case of having only $W=1$ and $W=-1$ depending on the sign of $\Delta_{\downarrow\downarrow}$ at $\theta=0$, one can induce topological phases with higher winding number and twist the boundaries. A variation of the second nearest hopping amplitude can bring to a complete reconstruction of the topological character. Indeed, as demonstrated in Fig.~\ref{fig:pwave-fm-phasedeltauudd-h2mu2t113}, the phase space is now marked by regions of $W=0,\pm 2$ at $\theta=0$. The change in the orientation of the magnetic field allows for the creation of regions with $W=\pm 4$ inside the $W=\pm 2$ domains or at the boundary between topological phases with $W=0$ and $W=\pm 2$.

\subsection{Topological phase transitions and marginal topological behaviour}

Under certain conditions the $p$-wave superconductor in the presence of a Zeeman field or a uniform exchange field can be gapless for an extended range of parameters.  However, this gaplessness can be removed by suitably perturbing the system in a way which opens the gap while preserving the symmetry properties.
We are interested in finding such possibilities and we start considering the case of a $p$-wave superconductor. 
Since the winding number determines the topology of the Hamiltonian, it is expected that whenever the winding number changes the energy gap has to close. 
One way to ensure that the determinant 
is zero, setting out the winding number, is by enforcing the imaginary part to be vanishing. Then, we only have to focus on  the real part.
Computing the determinant of $A_k$ (see Eq.\eqref{eq:ak-fm-pwave}), we get 
\begin{align*}
    \real|A_k| &= h_z^2 + h_y^2 - 4\sin^2(k)|\Delta_{\uparrow\uparrow}\Delta_{\downarrow\downarrow}| - \epsilon_k^2\\
    \imaginary |A_k| &= 2\sin(k)\left[h_z(\Delta_{\uparrow\uparrow} + \Delta_{\downarrow\downarrow}) + \epsilon_k(\Delta_{\uparrow\uparrow} - \Delta_{\downarrow\downarrow})\right] \\ 
    &= 4\sin(k)\left[h_z d_y + \imag\epsilon_k  d_x\right]
\end{align*}
Looking at the equations above, we see that one can ensure that the imaginary part is zero by setting $h_z = 0$ and $d_x = \frac{\Delta_{\uparrow\uparrow}-\Delta_{\downarrow\downarrow}}{2\imag} = 0$. This condition simultaneously indicates us the path to open the gap by externally perturbing the system. Indeed, one can either apply an out of plane 
magnetic field $h_z$, or alternatively alter the balance among the spin dependent superconducting order parameters. The first method is probably the simplest one from an experimental perspective.
As explained previously, the superconducting order parameters are not able to change the parity of the winding number at all. This means that if the parity is odd, and the gap at time reversal invariant momenta ($0,\pi$) is far from being zero, we are guaranteed that the system is in a marginal topological phase. Looking at Fig.~\ref{fig:pwave-fm-phasedeltauudd-h2mu2t11-2}, we see that this is the case for the line $\Delta_{\uparrow\uparrow}=\Delta_{\downarrow\downarrow}$ when $\theta=\pi/2$.  
Thus, we can state that any system along that line behaves as a marginal topological superconductor.
One way to guarantee that the system enters a topological state when we open the gap is by ensuring that the topological constant 
\begin{align}
    \mathcal{M} &= \sgn \real |A_0|\cdot \real |A_{\pi}|,\label{eq:fm:marg-topo-const} \nonumber \\
    &= \sgn \left[(-\epsilon_0^2 + |\bm h|^2)(-\epsilon_\pi^2 + |\bm h|^2)\right]
\end{align}
is negative. Since $\imaginary |A_k|$ is odd in $k$, whereas $\real |A_k|$ is even, the system is forced to have an odd winding number (except in the gapless case where it is ill defined). Thus, as long as the perturbation does not change the sign of $|A_k|$ at either time reversal invariant points, the gap opening immediately brings the system into a topological phase. 



\begin{figure}[tbh]
\begin{center}
   {\large (a)}\includegraphics[width=0.9\columnwidth]{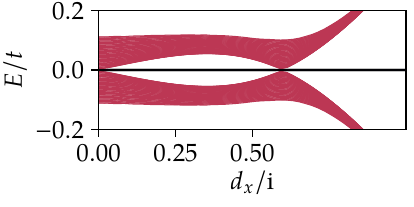}\\ \vspace{0.4cm}
   {\large (b)}\includegraphics[width=0.9\columnwidth]{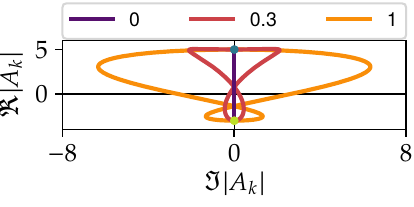}
\end{center}
    \caption[Marginal topological superconductivity by varying $d_x$.]{Marginal topological superconductivity by varying $d_x$. (a) Energy bands for open boundary system of 1000 lattice sites as function of the perturbation $d_x$. (b) Winding curves for different values of $d_x/\imag$ \cite{note}. 
    The parameters used relative to $t$ was $t' = t'' = 0$, $\mu=1$, $h_y=2$, $h_z=0$, $d_y=1.3$ and $d_x$ as given in the figures.}
    \label{fig:fm-marginal:openboundary-fm}

\end{figure}

Instead of applying an out-of-plane magnetic field, we can change the OP by adding a $d_x$ component. This is experimentally harder to do as it enters into the way the superconductor reorganizes the order parameter in response to an external field. However, it is interesting to investigate how this path allows us to drive the topological phase transition. In the following we use the same values as in the last example, but we perturb the order parameter instead of the out-of-plane magnetic field. We hypothesise that if this perturbation is able to open the gap, then the system will be topological with an odd winding number. 
We already know that adding a $d_x$ component will add an imaginary part to the determinant $|A_k|$. This is clearly shown in Fig.~\ref{fig:fm-marginal:openboundary-fm}(b) where the winding curve is initially flat against the real axis, but as soon as the perturbation is applied the imaginary part becomes non-zero (red and orange curves). As seen in Fig.~\ref{fig:fm-marginal:openboundary-fm}(a),
the bulk band gap (red) is opened when the $d_x$ component is increased from 0. In addition, the system goes into a topological phase, as indicated by the presence of the zero energy state for a range of values of $d_x$. When the $d_x$ component reaches approximately $0.6t$, the gap closes again at given $k$ distinct from the time reversal invariant points 0,$\pi$. However, although the gap closes, the system is still in a topological phase when the gap reopens, with a winding number of $|W|=1$, except that the sign changed. This band gap closing is due to the point where the curve (Fig.~\ref{fig:fm-marginal:openboundary-fm}(b)) 
intersects itself going through the origin. In this process, the winding direction of the curve changes.

\begin{figure*}[tbh]
    \begin{center}
{\large(a)}\includegraphics[width=0.4\textwidth]{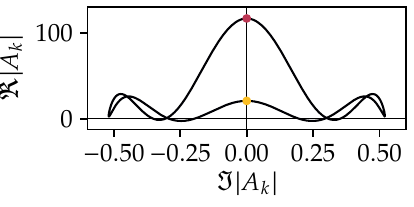}\hspace{0.65cm}
{\large(b)}\includegraphics[width=0.4\textwidth]{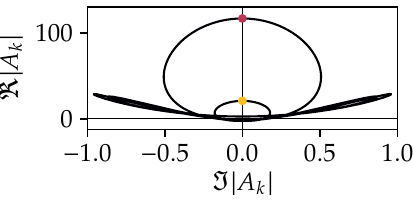}\\ \vspace{0.4cm}
\includegraphics[width=0.82\textwidth]{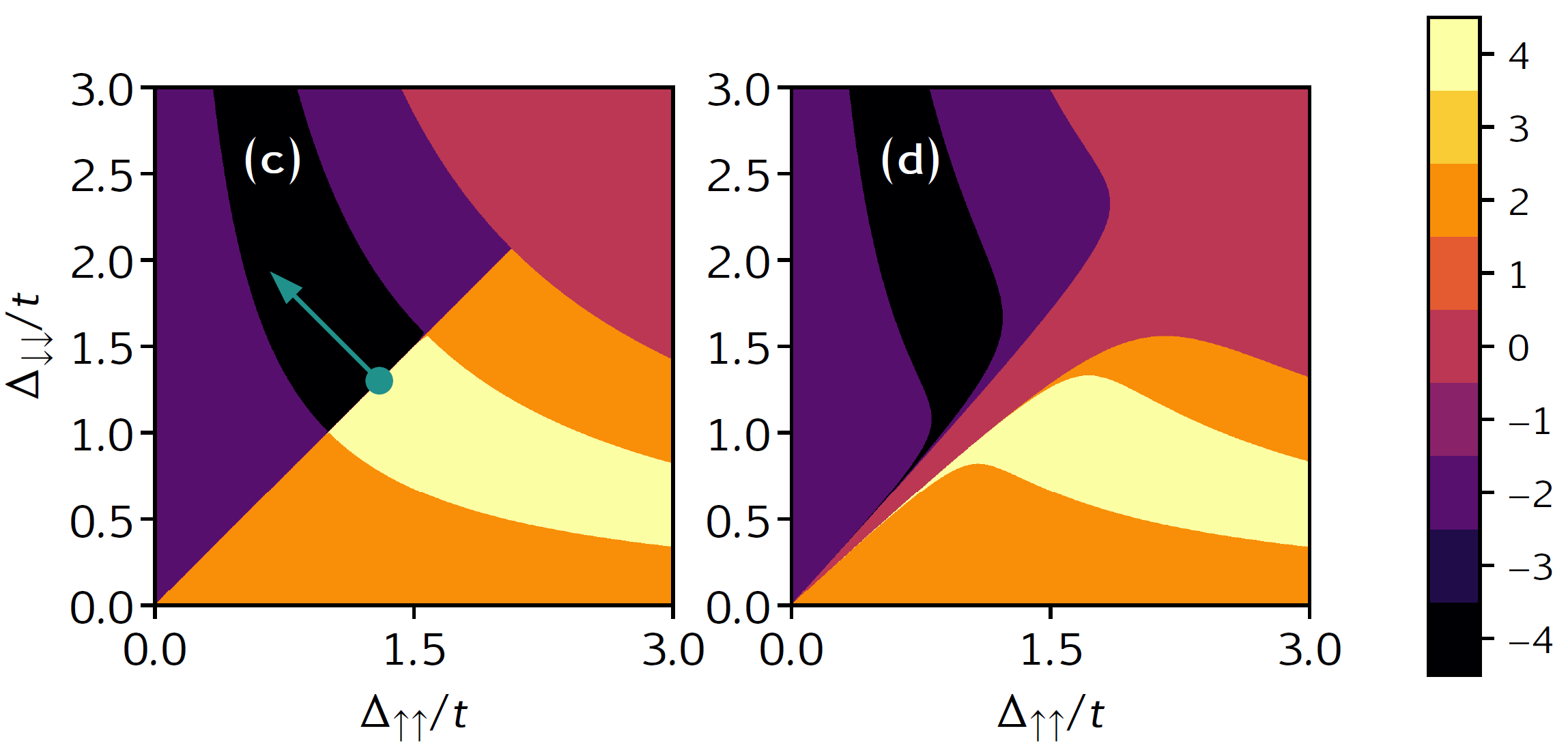}
    \caption[The superconductor can be marginal topological in one variable and trivial in another.]{(a) Plot the winding curve for $h_z=0.1t$ and $d_x=0$ and (b) for $d_x/\imag=-0.05t$ with $h_z=0$. {In both plots we have $d_y=1.3t$.}  (c-d) Phase diagram for $h_z=0$ and $h_z=0.1t$, respectively,  where the colors indicate the winding number (see colorbar). The other parameters for all the panels are $t' = t$, $t'' = 3t$, $\mu=1.0t$, $h_y=2 t$. {The green dot in panel (c)} shows the point in phase space that is perturbed, lying between two topologically nontrivial phases, and the arrow points in the direction of $d_x$ change. The superconductor can be marginal topological in one variable and trivial in another. When $\mathcal{M}$ is positive, the marginal topological nature of the system depends on the form the parametric curve $|A_k|$ takes when the system is perturbed. Here, the system is marginally topological in one variable ($d_x$) and switched between winding numbers 4 and -4, but is trivial in another ($h_z$). The reason for the trivial nature of the perturbation $h_z$ can be seen in the transition from (c) to (d), where the trivial area extends into the marginal topological area in the phase diagram. }
    \label{fig:fm:marginal-wind-and-phase}
    \end{center}
\end{figure*}

In Fig.~\ref{fig:fm:marginal-wind-and-phase} we show a case where the system can be marginally topological when perturbed in one variable, but be marginally trivial when perturbed in another. The reason for this is that   $\mathcal{M}$, given by Eq.~\eqref{eq:fm:marg-topo-const}, is positive in the case considered, which means that the winding number can only be even as long as the gap stays open at time reversal invariant $k$, i.e. $k=-k$. The evenness of the winding number does include zero as a possibility, which is the actual winding number that is achieved when the gap is opened by applying an out of plane magnetic field, $h_z$ (going from (c) to (d) in Fig.~\ref{fig:fm:marginal-wind-and-phase}). This is in contrast to the case when the gap is opened by adding a non-zero $d_x$ component (arrow in Fig.~\ref{fig:fm:marginal-wind-and-phase}(c)), which achieves a winding of $|W| = 4$, depending on the sign of $d_x$.
This difference in the topological phase in the perturbations can be explained by looking at paths that $|A_k|$ takes in the complex plane in each case. As seen in Fig.~\ref{fig:fm:marginal-wind-and-phase}(a-b) 
the shapes of the curves are qualitatively different in each of the cases, which is due to a difference in the dependence of the imaginary part on the perturbations, whereas the effects of the perturbations on the real part are negligible in the limit of small perturbations.

\begin{figure}[tbh]
    \centering
\includegraphics[width=0.98\columnwidth]{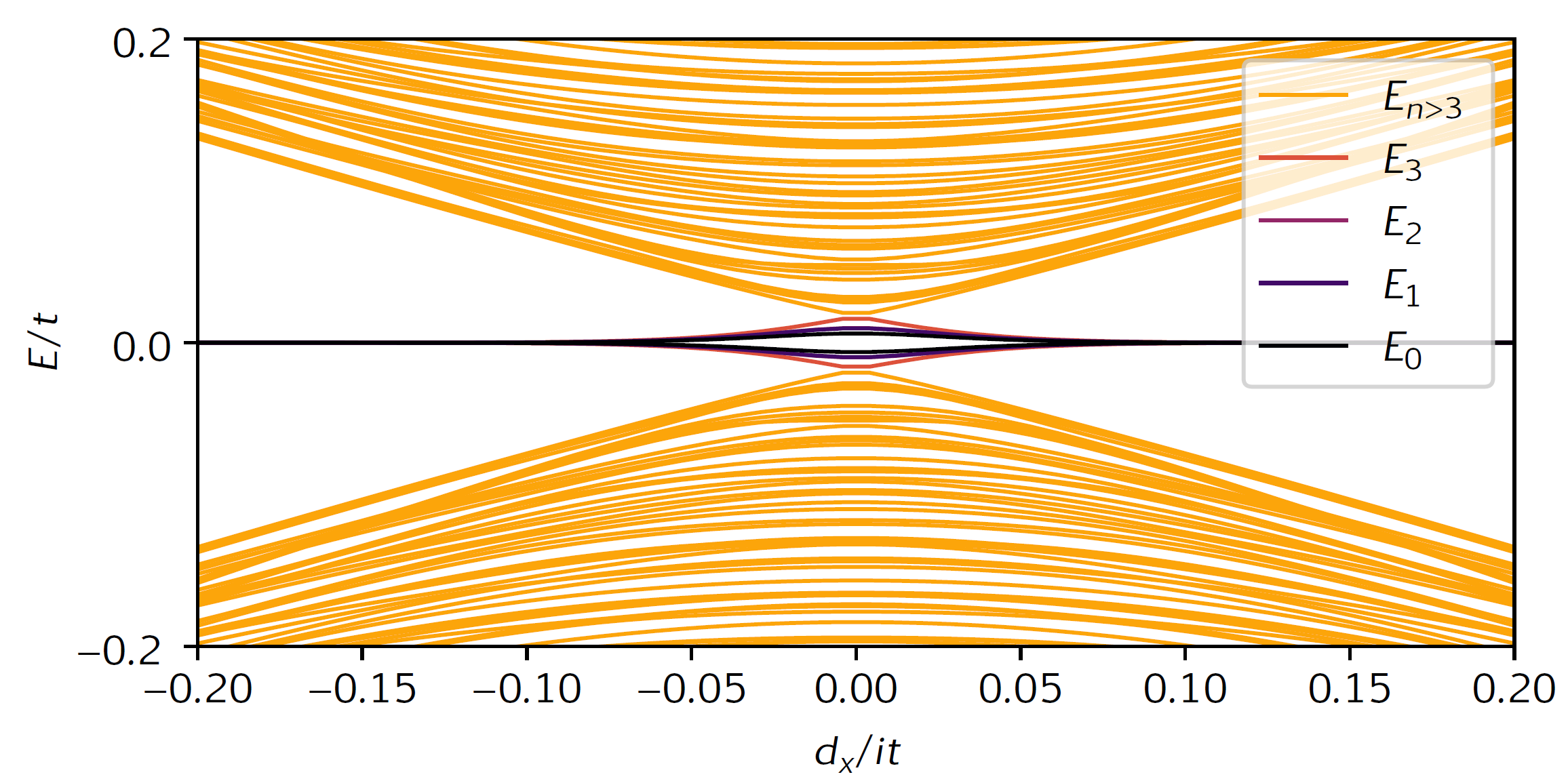}
    \caption{Open boundary energy bands as a function of $d_x$ for the parameters in 
    Fig.~\ref{fig:fm:marginal-wind-and-phase}, {namely  $t' = t$, $t'' = 3t$, $\mu=1.0t$, $h_y=2 t$ and $d_y=1.3t$ and for 1000 lattice sites. The small contributions seen close to $d_x=0$ are due to finite size effects.}}
    \label{subfig:fm:marg-open-dx}
\end{figure}

In Fig.~\ref{subfig:fm:marg-open-dx} the open boundary energy bands are shown as a function of the $d_x$ component, which is moving along the arrow in Fig. \ref{fig:fm:marginal-wind-and-phase}(c). 
This is located at a marginal point with   winding number  $|W|= 4$, meaning there are four {MBS} when we open the gap. Close to the marginal point, the {MBS} can be seen to interact significantly with each other leading to them gapping a little, and becoming extended states.

{From the observations above, we note the following: When the constant $\mathcal{M}$ is positive, the winding number is always even. This depends on the assumption that the determinant $|A_k| = |A_{-k}|^*$. When this is the case, we can cut the curve of $|A_k|$ in two by cutting at $k=0$ and $k=\pi$. Looking at each curve separately, it has to start and end at the same side of the origin on the real axis, and as a result has half-integer winding numbers each. This means that each half has an integer winding number, and since there are two, the winding number of the total curve can only be even. A similar geometrical argument applies to the case when $\mathcal{M} = -1$, where each of the points are on opposite sides of the origin on the real axis. 
We speculate whether the constant $\mathcal{M}$, which is computed at time reversal invariant momenta is linked to the product of the Pfaffians of the Hamiltonian at these two momenta or not. If this is the case, then one can expect that a negative $\mathcal{M}$ always gives a marginal topological state. }
The fact that a gapless superconductor with a positive $\mathcal{M}$ can be marginally topological and trivial at the same time implies that the topological nature of gapless systems is not as robust as when the parity of the winding number is odd, in which case \textit{any} gap opening (and symmetry conserving) perturbation pushed the system into a topological phase. 
When $\mathcal{M}$ is positive, we can only say for sure that the winding number is even, but it is harder to state whether the winding number is zero or non-zero. To do this, one must evaluate the exact dependence of the imaginary and real part of the determinant $|A_k|$.

\section{Topological phase transitions in superconductors with s-wave pairing and spatially modulated exchange field}
\label{sec:3}

When a s-wave superconductor is combined with a semiconductor having SOC in a Zeeman field, the resulting pairing, for some ranges of the parameters, is of $p$-wave type \cite{Kitaev_2001,Sato2010}, which can thus lead to MBS at the edge \cite{Alicea2011,Lutchyn2010}.
Alternatively, MBS can be obtained in systems consisting of an inhomogeneous exchange field impinged on the superconductor either through proximity or by magnetic atoms ~\cite{Choy2011,Nadjperge2013,Nakosai2013,Braunecker2013,Klinovaja2013,Vazifeh2013,Pientka2013,Poyhonen2014,Heimes2014,Brydon2015,Li2014,Heimes_2015,Xiao_2015,Andolina2017}.

In this section we analyse the topological phase diagram of an $s$-wave superconductor in an inhomogeneous and noncollinear exchange field with a focus on the emergence of a marginal topological phase. Since $s$-wave superconductors are more abundant than the intrinsic $p$-wave superconductors, the results from this section are relevant to deepen the investigation of another path to topological phase transitions in realistic heterostructures.

An extensively studied example of noncollinear exchange fields in $s$-wave superconductors is that of the helical exchange field \cite{Zlotnikov2021,RACHEL20251,Nakosai2013,Chen2015,Yang2016,Rex2019,Diaz2021,Garnier2019,Mascot2021,Maeland2023,Steffensen2022,Huang2023,Petrovic2021}, which is equivalent to a semiconductor with Rashba spin-orbit coupling (SOC) in a ferromagnetic field. These systems can be unitarily transformed to each other by performing a local spin-rotation, as shown in Appendix~\ref{app:B}. This type of system is an effective way of obtaining strong SOC fields which can be comparable to the electrons kinetic energy scale.
We chose to look at systems with translational symmetry, consisting of unit cells of a given number of lattice sites. Additionally, the exchange field vectors within the unit cell are in general independent of each other and lie in the same plane, thus not breaking the chiral symmetry.


The model we consider is given by the following tight-binding Hamiltonian 
\begin{equation}
    \begin{split}
        \ham =& 
        -t\sum_{\langle ij\rangle\sigma} c_{i\sigma}^\dagger c_{j\sigma} -\mu\sum_{i\sigma} c_{i\sigma}^\dagger c_{i\sigma}\\
        &-\sum_{iss'}c_{is}^\dagger \left(\bm h_i\cdot \bm \sigma_{ss'}\right) c_{is'}
        + \sum_{i}\left(\Delta_0 c_{i\uparrow}^\dagger c_{i\downarrow}^\dagger + \mathrm{h.c.}\right)\,,
    \end{split}
    \label{eq:synt:hamiltonian}
\end{equation}
which describes a superconductor at chemical potential $\mu$ with nearest neighbour hopping strength $t$, in the presence of exchange field $h_i$ at lattice site $i$. The $s$-wave superconducting OP, $\Delta_0 \equiv \Delta_{\uparrow\downarrow}$, 
is assumed to have a constant given amplitude in the model. For now, we only assume that the orientations of the $h_i$ are in the $xy$-plane, in which case the model has a global chiral symmetry. The choice of the plane is however arbitrary because the pairing is in the spin-singlet channel, so the results can be applied to any plane in the spin space. The model has particle-hole symmetry, which means that it 
must have a time-reversal symmetry too as given by the combination of the particle-hole and chiral symmetry transformations. The symmetry operators are given by
\begin{align}
    \TRSF &= \tau_z\sigma_x\CC & \PHSF &= \tau_x\CC & \SHSF &= \tau_y\sigma_x.
    \label{eq:synt:symmetries}
\end{align}
These symmetry operators hold as long as the superconducting order parameter is real, which can be always achieved by a global $U(1)$ gauge transformation in absence of flux biases. All the symmetry operators in Eq. \eqref{eq:synt:symmetries} square to one. The system in question belongs to the topology class BDI, which can have a $\mathbb{Z}$ topological invariant.

Hence, one question which arises is if and how an {inversion symmetry breaking collinear field} can lead to a topological non-trivial phase.
Since such a system has chiral symmetry, the winding number is a good topological number to classify the system by. Indeed, we obtain a winding whenever the determinant of the off-diagonal matrix $A_k$ winds around the origin. To do so, both the imaginary and real parts must be non-zero for at least some values of $k$ and have positive and negative values as $k$ goes from 0 to $2\pi$.
If either the real part or the imaginary part is constant zero in the reduced first Brillouin zone, the system can potentially be a marginal topological superconductor if it has gapless excitations.
Thus, following the same strategy of the case for the $p$-wave superconductor in Sec.\ref{sec:sec2_pwave}, one way to analyse the system is to inspect the imaginary component of the determinant of the off-diagonal matrix $A_k$ of the Hamiltonian. The form the imaginary part takes depends on the size of the unit cell, and is increasingly more structured when the size of the unit cell grows. We therefore compute it for few representative cases to capture the main effects and we focus on unit cell sizes with a number of sites that does not exceed four. We consider 
the Hamiltonian as in Eq. \eqref{eq:synt:hamiltonian} 
where the magnetic configuration has a periodic pattern that repeats itself after a number of sites $N_{\text{UC}}$. 
That means that the system is completely defined by the parameters $\mu$, $t$, $\Delta$ and $\bm h_1,\, \ldots, \bm h_{N_\mathrm{UC}}$.
The resulting expressions for the imaginary part of the off-diagonal determinants are $\imaginary |A_k| = 0$ for a unit cell with 
$N_\mathrm{UC} = 1$ and 2, while for 
$N_\mathrm{UC} = 3$ and 4 we have :
\begin{eqnarray}
     \imaginary |A_k| &=& -4\Delta t^3\sin(3k)\left|\bm h_1 \times \bm h_2 + \bm h_2 \times \bm h_3 + \bm h_3 \times \bm h_1 \right|, 
     \label{eq:marg:ak-n3}\\
    \imaginary |A_k| &=& 8\Delta\mu t^4\sin(4k)\left|\bm h_1 \times \bm h_2 + \bm h_2 \times \bm h_3 +\right. \nonumber \\
    &+& \left. \bm h_3 \times \bm h_4 + \bm h_4 \times \bm h_1 \right|,  
    \label{eq:marg:ak-n4}
\end{eqnarray}
respectively.
For a larger pitch of the magnetic pattern, when the unit cell size increases the expression for the imaginary part of the determinant of $A_k$ becomes more structured in a way that it cannot be expressed into a simple compact form (for instance, in addition to the cross products, terms like $h_i^x h_l^x h_k^x h_l^y$ appear). Thus, analytically solving for the imaginary part of the determinant of $A_k$ is not immediately affordable and gives little insight into the result. 
We observe then 
that systems with a unit cells of at most two lattice sites can never be topologically non-trivial since they are not able to have a non-zero winding number. Thus, ferromagnetic and antiferromagnetic systems are always topologically trivial. 
In the ferromagnetic case for example (unit cell of one lattice site), the gap is allowed to close when the magnetic field reaches $\sqrt{\Delta^2 + (2|t|-\mu)^2}$.
In contrast to the helical case, the gap will not open again after it first closes when $h$ is further increased. Instead, the Fermi points will move between $k=0$ and $k=\pi$, before it gaps out when it reaches a time reversal invariant momentum again.  

Looking at Eqs. \eqref{eq:marg:ak-n3}-\eqref{eq:marg:ak-n4}, we note that the imaginary part of $|A_k|$ is proportional to the area between the origin and the line segments connecting neighbouring magnetic fields (this does not generalise for  $N_\mathrm{UC} > 4$). A consequence of this is that whenever the field vectors $\mathbf{h}$ are collinear (i.e. they lie on any line in the $xy$-plane), the total area is zero, and {there is no net winding}. 
Based on these observations, we wondered if this could be generalized to bigger unit cells. Thus, we numerically computed the curves in the complex plane made by $|A_k|$ as $k$ traversed the 1BZ when the magnetic field vectors are
collinear with randomly selected amplitudes
and repeated this procedure for different number of unit cell sizes. 
We indeed obtained that the imaginary part of $|A_k|$ was always zero.
Thus, as first observation, 
we can state that an
$s$-wave superconductors with magnetic exchange vectors $\bm h_i$ which are collinear 
in the $h_xh_yh_z$-space will always be topologically trivial. 

The second general observation we make is that the dependence of the imaginary part of $|A_k|$ on the exchange field vectors is at most a first order polynomial in each component. That is, terms proportional to e.g. $h_1\,h_2\,h_3$ are allowed, but terms proportional to $h_1\,h_2^2$ are not allowed. This means that any time the imaginary part is purely zero, the winding number is an odd function in each perturbation of a field vector in close proximity. However, for larger perturbations, the real part of the determinant must also be considered, which may close the gap in time reversal invariant momenta asymmetrically in the amplitude of the perturbation. 

We will now consider different examples of magnetic textures , all of which exhibit gapless superconducting states, 
and single out the conditions 
under which this state can be converted into a topological or trivial phase.

\subsection{Example 1: unit cell of three lattice sites}

\begin{figure}[bt]
    \centering
    (a)\includegraphics[width=0.99\columnwidth]{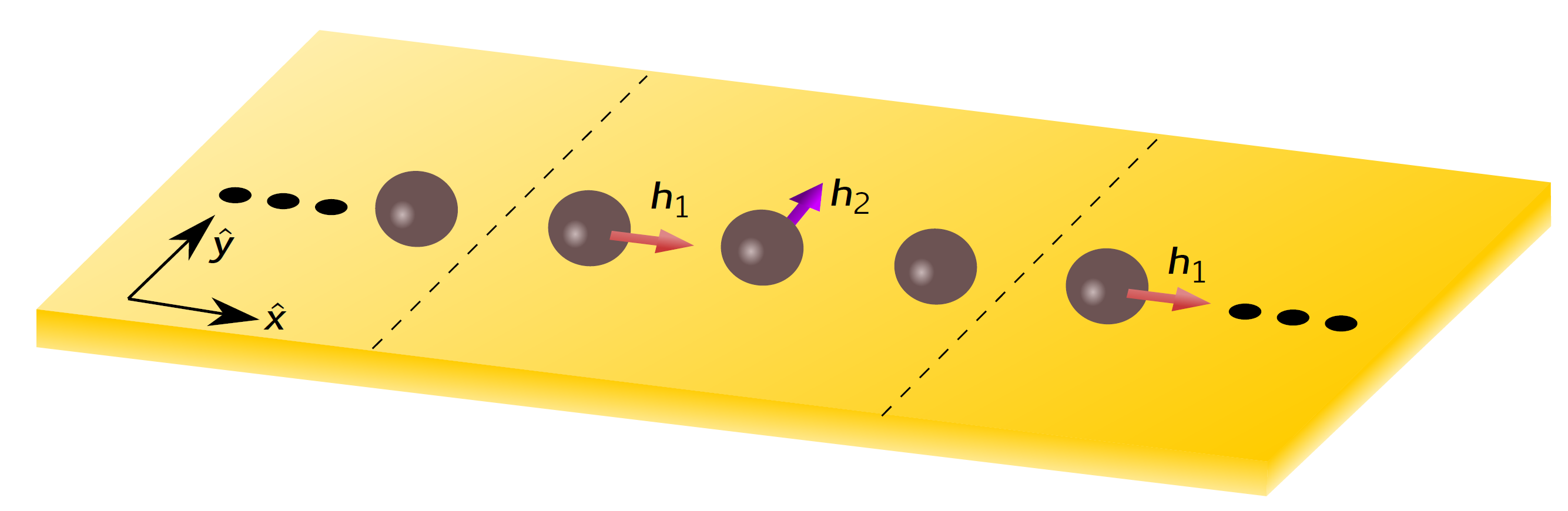}\\
    (b)\includegraphics[width=0.98\columnwidth]{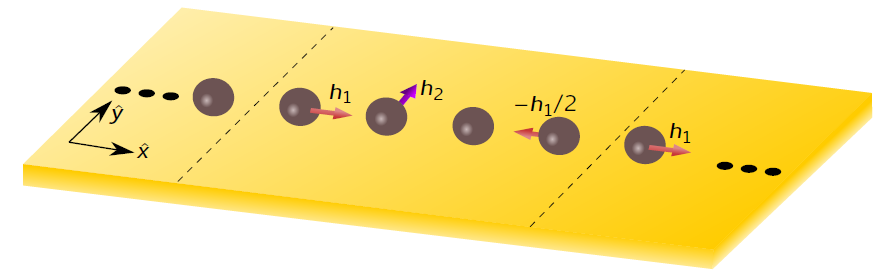}
    \caption{Schematic drawing of a superconducting chain with three (a) and four (b) sites in the unit cell. 
The arrows show the direction of the local exchange field, and the dashed lines indicate the limits of the unit cell, showing one lattice site from each neighbouring unit cell.}
    \label{fig:schematicfigexample-uc3}
\end{figure}

\begin{figure*}[bth]
    \begin{center}
    \includegraphics[]{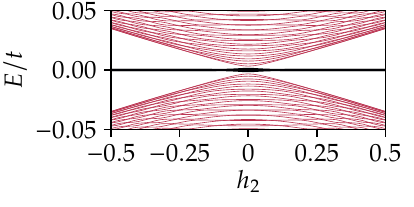}(a)
    \includegraphics[]{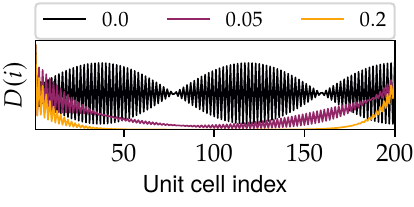} (b)\\ \vspace{0.8cm}
    \includegraphics[]{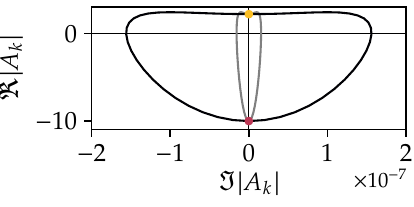}(c)
    \includegraphics[]{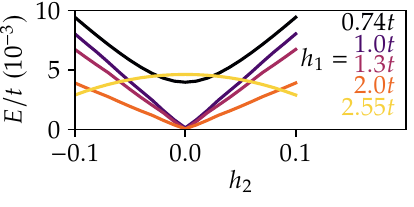}(d)
    \caption{
    \textbf{(a)} Energy bands  for an open boundary system as the perturbation $h_2$ is varied {for a system of 200 unit cells (i.e. 600 lattice sites). The energy bands} are shown in light (red) for the bulk bands, and black for the lowest lying energy. {For $h_2=0$ the system is gapless, and the small
contribution seen in the plot comes from  finite size effect}. \textbf{(b)} {The state with lowest energy extends through the whole system when it is gapless (black). As the gap is opened, the state becomes more localised at the edges (orange).} (c) Winding curve made by $|A_k|$ for $h_2= 10^{-8} t$ (gray) and $h_2= 10^{-7} t$ (black). (d) Bulk band gap as a function of $h_2$ for various values of $h_1$. The parameters for the plots are $\mu=-1.5t$, $\Delta = 0.3t$, $h_1=1.3t$ and $\bm h_2 = h_2\hat{\bm y}$, except for the panels that state otherwise.}\label{fig:marg-res}
    \end{center}
\end{figure*}

As a first instructive example we will consider a system with a unit cell of three lattice sites (see Fig.\ref{fig:schematicfigexample-uc3}(a)) where the first lattice site has an exchange field of $\bm h_1 = h_1\hat{\bm x}$, {the second $\bm h_2 = h_2\hat{\bm y}$, while on the third site it is ${h}_3=0$}.
Such systems can be realised, for instance, by  having different chemical elements in the unit cell with thus inequivalent magnetic moments.
Looking at \eqref{eq:marg:ak-n3}, we deduce that when $h_2 = 0$, the imaginary part of the determinant $|A_k|$ is zero, thus the system is either gapless or gapful and topologically trivial as $h_1$ varies. However, for non-zero $h_2$ the three points projected by $\bm h_1$, $\bm h_2$ and $\bm h_3$ create a triangle with a net area 
and thus the imaginary part of $|A_k|$ is non-zero. Indeed, the imaginary part is given by $-4\Delta_0 h_1 h_2 t^3 \sin(3k)$ and it is bilinear in $h_1$ and $h_2$. 
Keeping $h_1$ constant, we plot the energy bands for an \textit{open boundary} system as $h_2$ is varied in Fig.~\ref{fig:marg-res}(a). The figure shows that our predictions about the behaviour of the bulk system matches with the open boundary system. When the perturbation, $h_2$, is zero, the bulk band gap is closed except a small contribution from the finite size effect, which is expected to diminish as the length of the chain is increased. As $h_2$ is increased (or decreased) from 0, the gap opens but the lowest lying energy state remains at 0 energy and it is protected by a topological winding number of $\pm 1$, thus represents a Majorana mode. 

We computed the local density of the lowest energy state for three values of $h_2$ as it was increased from 0 to demonstrate how the localization of the MBS at the edge of the system is modified as the gap is opened. This is shown in Fig.~\ref{fig:marg-res}(b): 
when $h_2=0$, and hence the remaining magnetic exchange fields are collinear and the gap is closed, the Majorana mode extends through the whole superconductor, which allows the two Majorana modes to hybridize and get split energy. It further shows that a magnetic field strength of only $h_2=0.05t$ (purple line),  which is small compared to $\Delta$ ($h_2 / \Delta \approx 0.17$) and even smaller compared to the magnetic field at the first site ($h_2/h_1 \approx 0.04$), is enough to considerably localise the Majorana mode at the edge in a 200 unit cell system. As expected the increase $h_2$ leads to a further enhancement of the localization of the Majorana state at the edges of the chain corresponding to the sites at 0 and 200 (orange line).
In Fig.~\ref{fig:marg-res}(c) 
we display the winding curves for different values of $h_2$, showing how the imaginary part of the curve is considerably affected by small values of $h_2$, compared to the real part. 
The last figure, Fig.~\ref{fig:marg-res}(d),
shows how the band gap (scanned over the whole 1BZ) changes as $h_2$ varies. The lines are generated for different values of $h_1$, where the smallest and largest values are in the topological trivial regime, while the three middle values are in the marginal topological regime. 
As illustrated in the figure, when $h_2 = 0$, the gap stays closed for an extended range of $h_1$, before it reopens to a topological trivial phase. The last observation we make from the obtained results is that the rate at which the gap opens as a function of $|h_2|$ is sensitive to the size of $h_1$, as can be seen by the change of the slope of the lines. 

\begin{figure}[bth]
\centering
\includegraphics[width=1.\columnwidth]{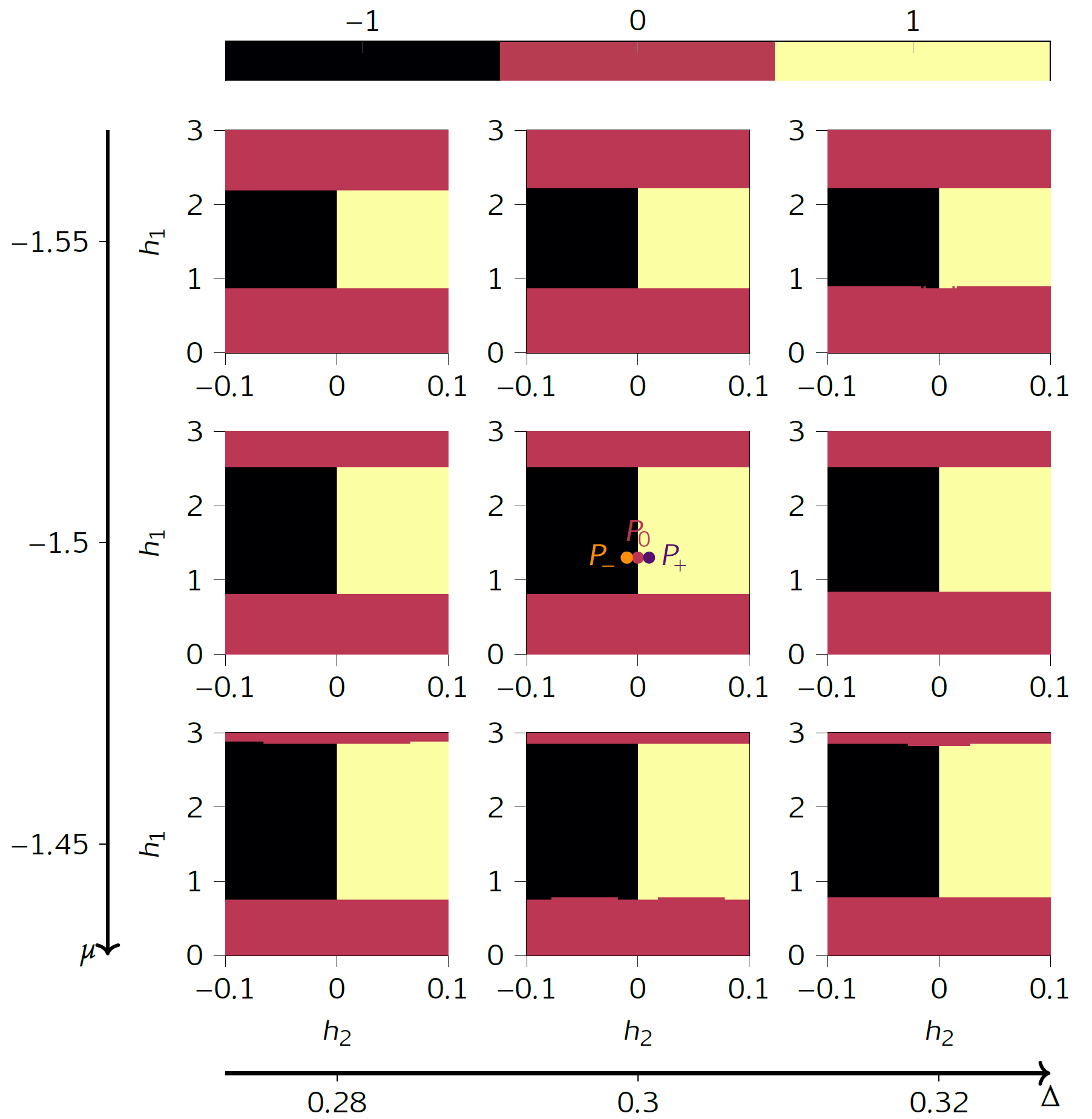}
\caption{Phase diagram for the three-atom-basis  system of example 1, which 
can be a marginal topological superconductor: If the system lies at a point in parameter space in between the black and yellow areas, any perturbation will bring the system into a topological phase, with the sign of the perturbation $h_2$ deciding the sign of the winding number. At the boundary between 
$W=-1$ and $W=+1$, the imaginary part of the determinant $|A_k|$ is zero for all $k$, with the Hamiltonian having two nodes at $k\neq -k$ points. The figure consists of nine plots showing the phase diagram in $h_2$-$h_1$-space 
arranged on a 3-by-3 grid, where the columns are parameterised by the superconducting OP, $\Delta$ and the rows 
by the chemical potential, $\mu$. 
{The winding curves of $|A_k|$ for the highlighed points in the central panel are shown in Fig.\ref{fig:non-coll:marginal-wind-3uc-h2}}
}
\label{fig:non-coll:phase-diagram-3uc}
\end{figure}

As was defined earlier, a marginal topological phase is described as a collection of points in parameter space where the Hamiltonian is gapless and any perturbation of the system is likely to bring it into a topological phase. 
{The case considered in this subsection is no different}. All of the parameters as given in Fig.~\ref{fig:marg-res} 
were perturbed, including $h_2$, and the resulting phase diagram is shown in Fig.~\ref{fig:non-coll:phase-diagram-3uc}. 
The phase diagram is antisymmetric in $h_2$, with a clear line dividing to topological phases with different winding number. 
This line in $h_2h_1$-space is stable against perturbations in the other parameters and is embedded inbetween topological phases. That is, any system lying on the line will become topological when perturbed, provided that $h_2$ is non-zero. 
{As we have demonstrated, marginal topological systems may have an imaginary part which depends on only a small number of parameters. Here, that parameter is $h_2$. }

\begin{figure}[tbh]
    \centering
    \includegraphics[width=1.\columnwidth]{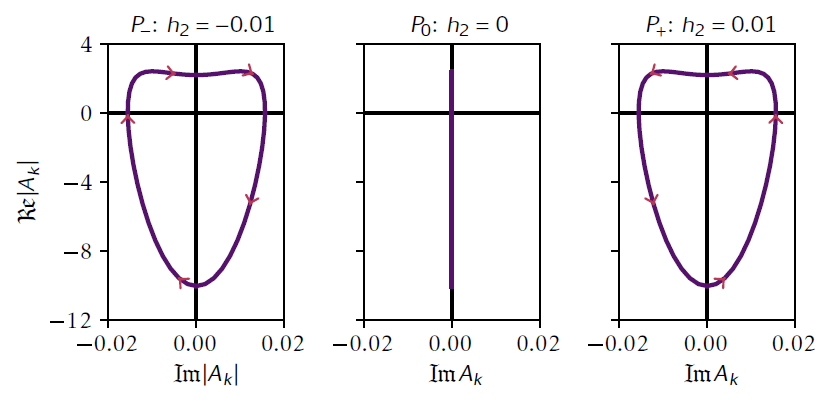}
    \caption{Winding curves for different values of the parameter $h_2$ shown as the points $P_{-}$, $P_0$ and $P_+$ in Fig.~\ref{fig:non-coll:phase-diagram-3uc}.}
    \label{fig:non-coll:marginal-wind-3uc-h2}
\end{figure}


\subsection{Example 2: unit cell of four lattice sites}

We have 
demonstrated that certain gapless superconducting systems can be easily driven into a topological phase by just perturbing the amplitude of the magnetic exchange solely at one site in the unit cell, thereby opening the gap. However, we propose a further requirement for when the system is a marginal topological superconductor and when it is a trivial gapless system. 
We argue that the specific class of systems considered will always have an imaginary part of $|A_k|$ which is proportional to $\sin{(q\,k)}$ (with $q=1/N_{\mathrm{UC}}$ being the inverse of the number of sites in the unit cell). In other words, as $k$ traverses the Brillouin zone, the imaginary part of the determinant of $A_k$  oscillates once. From this basic observation, the curve winds \textit{if and only if} the number
\begin{equation}
    \mathcal{M} = \sgn \real |A_0|\cdot \real |A_{\pi/q}|
\end{equation}
is $-1$, i.e. the real part has opposite sign at $k=0$ and $qk=\pi$. 
The system described in the previous case satisfies this criterion. However, to demonstrate the opposite case, in which $\mathcal{M}=+1$, we consider a system with a unit cell consisting of four lattice sites, and where the exchange field is given by $\bm h_1 = h_1\hat{\bm x}$, $\bm h_2 = h_2\hat{\bm y}$, $\bm h_3 = -h_1\hat{\bm x}/2$  and $\bm h_4 = 0$. This physical configuration is depicted in Fig.~\ref{fig:schematicfigexample-uc3}(b). 
\begin{figure}[tb]
    \begin{center}
    (a)\includegraphics[width=0.8\columnwidth]{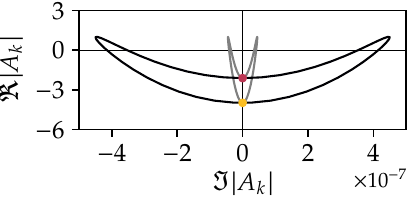}\\
    (b)\includegraphics[width=0.8\columnwidth]{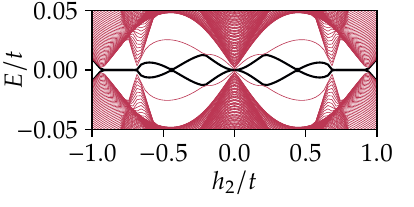}\\
    (c)\includegraphics[width=0.8\columnwidth]{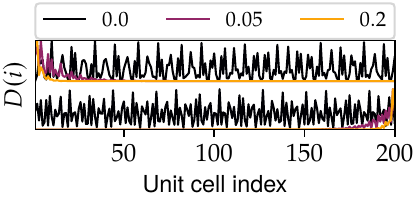}
\caption{Analysis of a system consisting of unit cells with  four lattice sites which is gapless, yet not marginally topological.
The parameters used are $\mu = -0.5t$, $\Delta = 0.3t$, $h_1 = 2.5t$ and $N_\mathrm{UC} = 200$. (a) The winding curve is drawn for $h_2/t = 0.005$ (gray curve) and $h_2/t = 0.05$ (black curve) demonstrating how the {\it banana} shape curve does not enclose the origin. The two dots indicate the point on the black curve where $k=0$ (red) and $qk=\pi$ (yellow), and show how they are located on the same side on the real axis, giving $\mathcal{M} = +1$. (b) The energy bands for an open system are computed as a function of $h_2$, and shows that as the gap is initially opened, the lowest lying energies have non-zero energy. When $h_2/2$ reaches $\approx 0.65$, the gap closes again, this time at a time reversal invariant momenum, thus leading to a non-trivial winding number. (c) Local density of state for the first  (bottom panel) and second (top panel) excited states are drawn for different values of the perturbation, $h_2/2$.  The system has two edge states for $h_2\neq 0$, both of which have finite energy.}
    \label{fig:marg-res-trivial}
    \end{center}
\end{figure}
The curve that $|A_k|$ takes in the complex plane (for a large negative amplitude of $h_2$) is shown in Fig.~\ref{fig:marg-res-trivial}(a). 
As can be seen, the real part of $|A_k|$ at $k=0$ (red point) and $qk=\pi$ (yellow point) have the same sign. This leads to the curve not enclosing the origin, which means that the system is \textit{not} a marginal topological superconductor. Thus, if the system is perturbed in a way that opens the gap, there will not be any Majorana modes at the boundaries.
As Fig.~\ref{fig:marg-res-trivial}(c) 
shows, there are two excited states that localises at the edges when the gap is opened, but as Fig.~\ref{fig:marg-res-trivial}(b) 
shows, these states have non-zero energies. 
This is in contrast to the topological phase where there are two Majorana edge states.

As a side note, as the perturbation $h_2$ is further increased to around $h_2 = 0.65t$, the system does close the gap at a time reversal invariant momentum, which brings the system into a topological phase. This can be seen as the regions in Fig.~\ref{fig:marg-res-trivial}(b)  
where the black band is flat. In these regions, the winding number is $\pm 1$. However, this is not considered as a marginal topological region.

\begin{figure}[t]
    \centering
    (a)\includegraphics[width=0.75\columnwidth]{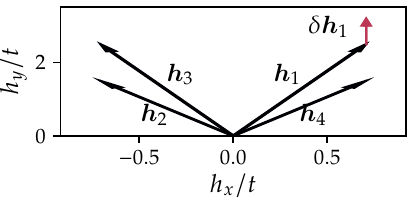}\vspace{0.5cm}
(b)\includegraphics[width=0.99\columnwidth]{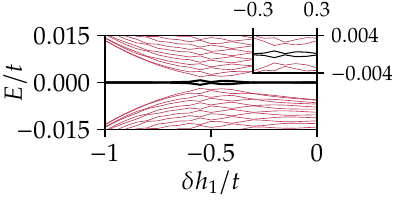}
\caption{
(a) Schematic illustration of the magnetic pattern.
The red arrow is the perturbation applied to the system. (b) Energy bands as functions of the perturbation $\delta \bm h_1$. The lowest lying energy is shown in black, while the bulk energy bands are shown in red. The inset shows the same case, but zoomed in. The parameters used are $\mu=-1.5t$, $\Delta=0.3t$ and $N_\mathrm{UC} = 400$. Since the hybridization between {Majorana modes} at opposite edges decrease exponentially with system size \cite{Kitaev_2001}, the system size was increased to 400 unit cells to show the effects more clearly. 
}
\label{fig:number-8}
\end{figure}

In the last physical case of this section, we demonstrate that there exist other patterns than collinear patterns that are also marginally topological. The case we consider has a magnetic texture given in Fig.~\ref{fig:number-8}(a). 
This pattern is parameterised by $h_j/t = \cos(j\pi/2-\pi/4)\hat{\bm x} + (0.5\sin(j\pi/2-\pi/4) + 2)\hat{\bm y}$. 
One of the points is perturbed, as shown by the red arrow. The energy bands for the open boundary system are shown in Fig.~\ref{fig:number-8}(b),  
and the flat black band indicates that there is a MBS present at the boundaries when the first lattice site in the unit cell is perturbed along the direction of the red arrow, shown in Fig.~\ref{fig:number-8}(a).
In this system, the band gap opened slower than the previous cases, hence we had to increase the size of the finite system to 400 unit cells due to finite size effects. 
As displayed in the inset of Fig.~\ref{fig:number-8}(b),
when the band gap is only slightly opened, the {Majorana modes} at the ends are still hybridizing considerably, which leads to them gaining energy at first. As the energy gap is opened further by the perturbation, the interaction between the {edge states} 
diminishes due to their
increased inability
to propagate through the bulk. 

Thus as the gap increases, the energy of the Majorana modes approach zero. 
Although this system can easily be brought into a topological phase by a very small perturbation, we would still argue that it is not a marginal superconductor based on our definition of such systems. 


In the examples considered here, the functional form of the imaginary part of $|A_k|$ is 
odd in the perturbation. As a result, when the perturbation is applied with a negative sign, one obtains a topologically different phase. Both phases have the same amplitude of the winding number, but have opposite signs. The implications of this is that the compatibility of two domains with the same type of perturbation depends on the direction of the perturbation. 
If we for example take the case considered with three sites in the unit cell, as discussed above, and set the perturbation to $h_2\hat{\bm y}$ in the first domain, and couple this domain to a second domain with the perturbation $-h_2\hat{\bm y}$, two MBSs will appear at the domain wall between them. This is a result of the bulk-boundary condition, 
stating that the topological properties of a bulk system determine the existence of robust edge or surface states at its boundaries. Hence, two bulk systems with different topological invariants cannot be continuously deformed into each other without closing the energy gap, i.e., specifically for our case,
a clockwise oriented curve can not be transformed into a counterclockwise oriented curve without first squashing it and intersecting itself (see for instance Fig.\ref{fig:non-coll:marginal-wind-3uc-h2}), thereby forcing the gap to be closed somewhere. This can potentially be a path to easy control and 
transport  MBSs in practical systems as only small perturbations are needed to change the topology of regions of the device. However, the smaller the device is, the larger the perturbation needs to be since small gaps means that the MBS extends far into the bulk and will interact increasingly with small device sizes.


\section{Conclusion}

We analysed different types of superconducting systems, with a focus on topological transitions which can be achieved with small perturbations, that we indicated as a marginal topological 
superconductors. These
can potentially be important tools in devices where control, creation and annihilation of MBSs is required, as only minor changes in the parameters of the system can lead to localization of the MBSs at the edges. 
The criteria for model parameters that lead to marginal topological behavior can be easily identified by looking for cases where the trajectory associated with the winding number has a purely real amplitude.
Such a realness criterion of the winding curves 
relies on only a few parameters
in the examined systems, 
in a way that realness is achieved when these parameters are set to zero.
Searching for a vanishing amplitude of a polynomial functional can be generally satisfied and often for a large set of parameters, thus making the marginal topological phase in principle accessible for materials systems and heterostructures that are experimentally feasible. 
In  p-wave superconducting systems the realness criteria relies on the $d_x$ order parameter and the presence of a $h_z$ component of the applied field or of the ferromagnetic exchange field. 
From an experimental point of view, driving the marginal topological phase of p-wave superconductors might be most easily done by controlling the $h_z$ component of the exchange field, which in the case of an external field is as easy as rotating either the source of the magnetic field around the specimen or rotating the specimen itself.

The s-wave superconductor 
with spatially modulated exchange field
does only depend on one superconducting order parameter, which is in principle more feasible to control in contrast to the unconventional p-wave order parameter. In this framework, to realize a marginal topological phase 
one need a system with a unit cell of at least three atoms where the exchange field vectors are different enough, and the gap opening perturbations needs to be local in the unit cells, only affecting the exchange field at a few of the sites or affecting the sites differently.
If the magnetic unit cell consists of different atoms for example, we speculate that one can use short laser pulses to change the magnetic moment at each lattice site individually by tuning the laser pulse frequency accordingly \textcite{Kirilyuk2010}.

As demonstrated in Fig.~\ref{fig:marg-res-trivial}, gaplessness is not a sufficient requirement alone to have a marginal topological state. In the case where only nearest neighbour hopping is significant and the winding number can only be $-1$, $0$ or $+1$, the number of nodes of in the Brillouin zone are important. If the number of nodes in each half is even, the system is trivial. On the other hand, if it is odd, the system is guaranteed to be marginal topological. Indeed, this applies to all one dimensional systems with a  winding number, since the sign difference means that the winding number can only be odd, which excludes the trivial value of 0.
We also showed that the system can be marginal topological with an even number of nodes in each half of the Brillouin zone if there are next nearest and third nearest neighbour hopping in the system. Thus when the gap is opened multiple Majorana bound states can be directly generated at the edges of the superconducting nanostructure.

{It is important to note that the utilization of internal spin degrees of freedom is not a necessary prerequisite for a marginal topological superconductor. In this context, one can also consider other internal degrees of freedom, such as sublattice \cite{Leykam2018,Mizogichi2019,Shahbazi2023}, or orbital degrees of freedom \cite{Jo2024,Mercaldo2020,AQT_23,Z4TSC_2024}.
Specifically, it is, for instance, instructive to consider the sublattice degree of freedom. For this physical configuration, in order to achieve a marginal superconducting state, one has to assume a superconducting system characterized by a two-atom unit cell (A, B), that can be represented by a pseudospin-1/2 system. 
Within this framework, an electron situated at site A is represented by an effective pseudospin with up configuration, whereas an electron at site B corresponds to a pseudospin down. In this context, a spatial even-parity pairing featuring a spin-singlet with pseudospin-triplet configuration across different sites in the same sublattice is permitted by symmetry. It can be characterized by a pseudospin $d$-vector that mirrors the structure of spin triplet pairing with $d_x$ and $d_y$ components, if the pairing on for the electron on the A sites is different from that on the B sites. Drawing an analogy, the electron hopping -- both with and without inversion symmetry -- between the A and B sites simulates a magnetic field with varying orientations in the $xy$ pseudospin plane. Meanwhile, the charge imbalance between the two sites results in an effective $h_z$ field within the sublattice pseudospin framework.
The sublattice conserving hopping with first and second nearest neighbor has to be included for getting the high winding topological configurations.
Hence, the marginal topological superconductor can be obtained by suitably tuning the amplitude of the hopping and sublattice charge unbalance.
In the same fashion one can also design a marginal topological superconductor by exploiting the orbital degrees of freedom.}

\section*{Acknowledgments}
M.T.M acknowledges support from PNRR MUR project PE0000023-NQSTI (TOPQIN) and  partial support by the
Italian Ministry of Foreign Affairs and International
Cooperation PGR12351
(ULTRAQMAT). 
M.C. acknowledges partial financial support from PNRR MUR project PE0000023-NQSTI.
We acknowledge valuable discussions with Jacob Linder and Panagiotis Kotetes.

\appendix

\section{Simple illustration of marginal topological superconductor}
\label{app:A}

In this appendix we illustrate a general simple example of {\it marginal} topological superconductor.
A topological superconductor is a physical system whose properties are fundamentally characterized by topological invariants rather than local order parameters and its superconducting energy gap cannot be smoothly deformed to a trivial state without closing at some critical point.
Followingly, any time the gap closes, the system may change its topological character. 
An interesting {\it edge} case arises in one-dimensional gapless systems.
\begin{figure}[b]
\includegraphics[width=1.\columnwidth]{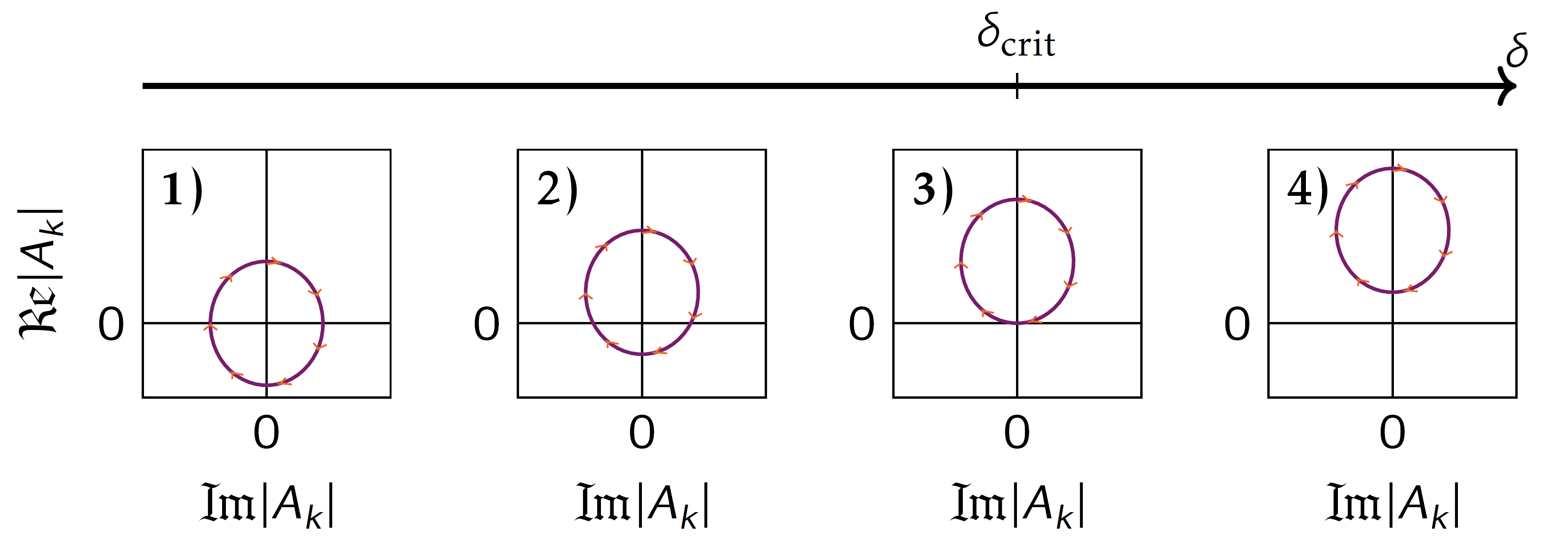}\\(a)
\\\vspace{0.35cm}
\includegraphics[width=0.99\columnwidth]{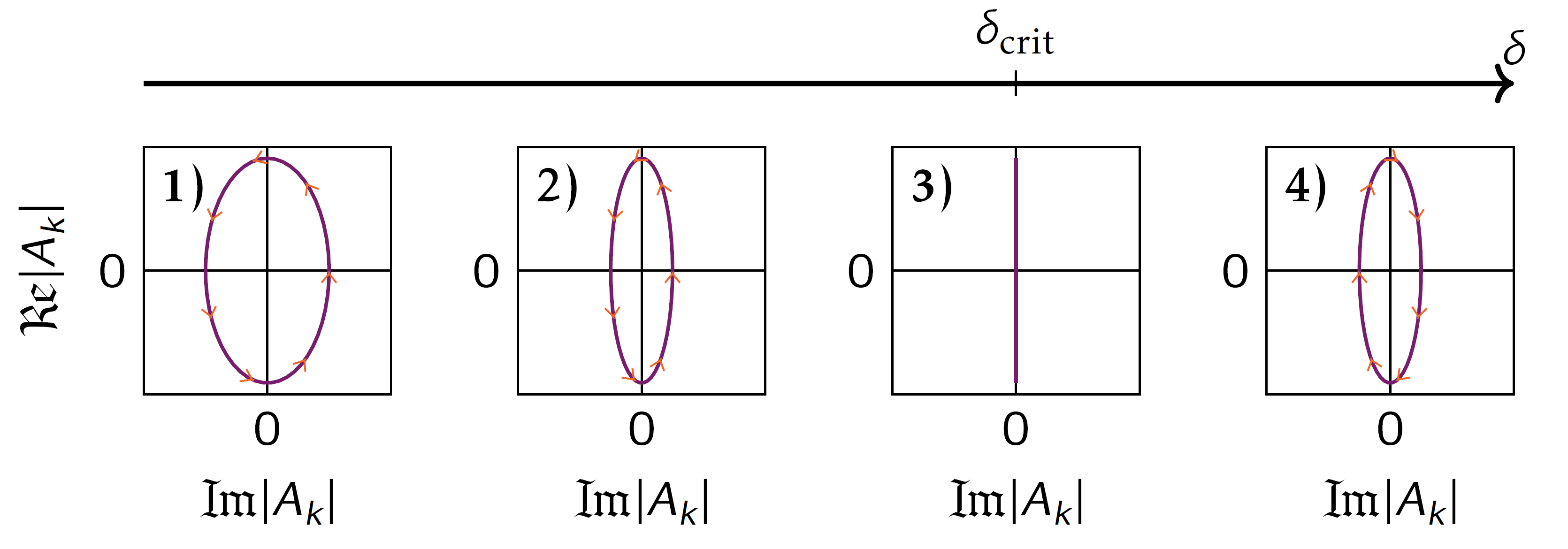}\\(b)
\caption{Schematic phase transtion for (a) {\it normal} and (b) {\it marginal} superconductors. The closed curves are the determinant $|A_k|$ parameterised by k in the first Brillouin zone, and the different panels show the evolution with respect to a parameter $\delta$.  For a normal topological phase transition (a) at the transition point, the winding number changes value. For a marginal topological superconductors (b) at the transition point the winding number changes sign.
}
\label{fig:A}
\end{figure}
We consider superconducting systems with a global chiral symmetry which has a topological invariant defined by the winding number $W$ \eqref{eq:winding-number-complex}, which is related to the determinant $|A_k|$ (cfr.\eqref{eq:block-off}). 
When 
$W\neq 0$ the  curve made by $|A_k|$, while varying $k$ in the 1BZ, in general encloses the origin of the complex plane $(\Re|A_k|,\Im|A_k|)$, see for instance Fig.\ref{fig:A}(a1). There are two ways in which the gap may close. The first, and most general, distinguishes topologically different phases from one another, presented in Fig.\ref{fig:A}(a). This occurs when a certain parameter, here labeled by $\delta$, reaches a critical value $\delta_{\mathrm{crit}}$. In this case, the gap closes and then reopens, accompanied by a change in the topological number.
In general, $\delta_{\mathrm{crit}}$ depends on all other parameters of the Hamiltonian, meaning that the gap closing typically requires fine-tuning.
The second way occurs when  the curve is a straight line including the origin of the complex plane (see Fig.\ref{fig:A}(b3)) and changes in the parameter $\delta$ drives to changes in the sign of $W$. This leads us to consider cases were the $|A_k|$ is purely real for all $k$ in the 1BZ. In Sec.\ref{sec:sec2_pwave} and \ref{sec:3} we showed some examples in which the realness-criterion relies only upon a subset of the parameters of the Hamiltonian and any change in the remaining parameters will not add any imaginary part to $|A_k|$. 

\section{Local spin rotation and helical magnetic field}
\label{app:B}

In Sec.\ref{sec:3} we introduced the Hamiltonian \eqref{eq:synt:hamiltonian} describing a s-wave superconductor in the presence of a Zeeman field. This can be transformed to a model with a collinear magnetic field by performing a local spin rotation around the $z$-axis to align all the spins along one axis -- in our case we choose the $x$-axis. If we define the angle the Zeeman field at lattice site $i$ makes in the $xy$-plane as $\lambda_i$, the local rotation at this lattice site is given by
\begin{equation}
    R_i = \exp[-\frac{\imag (\lambda_i) \sigma_z}{2}].
    \label{eq:synt:local-rotation-def}
\end{equation}
This transformation aligns all local Zeeman fields along the positive $x$-axis,
\begin{align*}
    R_i^\dagger \bm h_i\cdot \bm \sigma R_i  &= |h_i|\sigma_x.
\end{align*}
Thus, in this new basis, the magnetic field is that of a spin density wave (SDW). 
An effect of the local rotations is that whenever to adjacent Zeeman fields are initially oriented in different directions (independent on the actual size of the fields), the previous 
hopping parameter $t$ between the two lattice sites is transformed to a combination of 
opping and Rashba-type SOC as follows,
\begin{align*}
    t \to t R_{j}^\dagger R_{i} &= t\left(\cos(\delta\lambda_{ji}/2)\sigma_0 + \imag \sin(\lambda_{ji}/2)\sigma_z\right),
\end{align*}
where   \(\delta\lambda_{ji} = \lambda_j - \lambda_i\) is the angle difference.
The resulting Hamiltonian is equivalent to that of a singlet s-wave superconductor in a collinear SDW field with a {\it modulated} Rashba spin-orbit coupling:
\begin{equation}
    \begin{split}
        \ham_\mathrm{R} = &\sum_{\langle ij \rangle ss'} c_{js}^\dagger \left(t(j,i)\sigma_0 + \imag\alpha_R(j, i)\sigma_z\right)_{ss'} c_{is'} 
        - \mu\sum_{is} c_{is}^\dagger c_{is}\\
        &-\sum_{iss'}c_{is}^\dagger h_i (\sigma_x)_{ss'} c_{is'} 
        +\sum_{i}\left(\Delta c_{i\uparrow}^\dagger c_{i\downarrow}^\dagger + \mathrm{h.c.}\right)
    \end{split}
    \label{eq:synt:hamiltonian-after-rot}
\end{equation}


An important special case of the the model \eqref{eq:synt:hamiltonian} is when the local magnetic field is helical along the 1D wire, i.e. the fields lie on a circle with a constant angle difference between the adjacent sites. Thus, the rotated Hamiltonian describes a superconductor with ferromagnetic field of strength $h$ with hopping parameter $\Tilde{t}\cos(\delta\lambda/2)$ and Rashba SOC strength $\alpha_R=t\sin(\delta\lambda/2)$. 
This model has been already discussed \cite{Alicea2011,Oreg2010,Sato2010}, however in the context of this paper it is 
interesting to establish a connection between inhomogeneous magnetic fields and the collinear model with Rashba-like SOC.

The rotated system is translationally invariant, which implies that the momentum is a good quantum number. 
The Fourier transformed Hamiltonian is then
\begin{equation}
    H(k) = \begin{pmatrix}
        \epsilon_k\sigma_0 + \alpha_k\sigma_z - h\sigma_x & \imag|\Delta|\sigma_y \\
        -\imag|\Delta|\sigma_y & -\epsilon_k\sigma_0 + \alpha_k\sigma_z + h\sigma_x
    \end{pmatrix},
\end{equation}
where $\epsilon_k = -2\Tilde{t}\cos k - \mu$ and $\alpha_k = -2\alpha\sin(k)$ and the basis is the 
standard Nambu spinor.
Since the Hamiltonian has a chiral symmetry, we can block off-diagonalize the matrix as done in Sec.\ref{sec:sec2_pwave}, leading to the block matrix $A_k$
\begin{equation}
    A_k = \begin{pmatrix}
        \alpha_k + \imag |\Delta| -\epsilon_k & h \\ 
        h & -\alpha_k - \imag \Delta -\epsilon_k
    \end{pmatrix},
\end{equation}
whose determinant is given by
\begin{equation}
    |A_k| = \epsilon_k^2 - h^2 - \alpha_k^2 + \Delta^2 - 2\imag\alpha_k\Delta.
    \label{eq:helical:ak}
\end{equation}
The imaginary part can only be zero at $k=0$ and $k=\pi$. Thus the path in the complex plane can only touch the origin at the point given by $k=0$ or $k=\pi$. Consequentely, only the real part is necessary to determine when the winding number is non-zero. Inserting for $k=0$ and $k=\pi$, we get the following two independent equations for when the gap is allowed to close
\begin{align}
    (2\Tilde{t}+\mu)^2 + \Delta^2 = h^2 \quad (\text{gap closes at } k=0)\label{eq:helic:gapclose0}\\
    (2\Tilde{t}-\mu)^2 + \Delta^2 = h^2 \quad (\text{gap closes at } k=\pi)\label{eq:helic:gapclosepi}
\end{align}
It is interesting to note  that the topological phase diagram does not depend on the size of the Rashba SOC parameter, but only on the Zeeman field strength, the band width and the superconducting gap parameter. However, the size of the spin orbit coupling affects the size of the bulk band gap which in turn affects the robustness of the MBS against perturbations. 

%

\end{document}